\documentclass{article}
\usepackage{amssymb}
\usepackage{fleqn}
\usepackage{epsfig}
\usepackage{graphicx}
\usepackage{amsmath}

\def\be{\begin{equation}}
\def\ee{\end{equation}}
\def\bea{\begin{eqnarray}}
\def\eea{\end{eqnarray}}
\def\bq{\begin{quote}}
\def\eq{\end{quote}}

\def\gappeq{\mathrel{\rlap {\raise.5ex\hbox{$>$}}
{\lower.5ex\hbox{$\sim$}}}}
\def\lappeq{\mathrel{\rlap{\raise.5ex\hbox{$<$}}
{\lower.5ex\hbox{$\sim$}}}}
\def\Toprel#1\over#2{\mathrel{\mathop{#2}\limits^{#1}}}

\begin{document}
\pagestyle{empty}
\vspace*{35mm}

\begin{center}
\textbf{ The evaluation of loop amplitudes via differential equations } \\[0pt]

\vspace*{1.5cm}

\textbf{Ugo Aglietti}\footnote{e-mail address: Ugo.Aglietti@roma1.infn.it} \\[0pt]

\vspace{0.7cm}
Dipartimento di Fisica ``G. Marconi'',
Universit\'a di Roma ``La Sapienza'', \\
and INFN, Sezione di Roma, Piazzale Aldo Moro 2, 00185 Roma, Italia. \\[1pt]

\vspace*{1.5cm} 
\textbf{Abstract} \\[0pt]
\end{center}

The evaluation of loop amplitudes via differential equations 
and harmonic polylogarithms is discussed at an introductory level.
The method is based on evolution equations in the masses
or in the external kinematical invariants and on a proper 
choice of the basis of the trascendental functions.
The presentation is pedagogical and goes through specific 
one-loop and two-loop examples in order to illustrate the general 
elements and ideas.

\vspace*{6cm} \noindent %\rule[.1in]{16.5cm}{.002in}

\vfill\eject
%\pagestyle{empty}
%\clearpage\mbox{}\clearpage

\setcounter{page}{1} \pagestyle{plain}

% Beging of the text 

\section{Introduction}

The evaluation of radiative corrections in quantum field theory is 
notoriously a hard task and various methods have been proposed
in decades to accomplish it, such as Feynman parameters, dispersion 
relations, low-momentum expansions, etc.
In the past few years, a new method has been developed, 
which is based on $(i)$ the reduction of the amplitudes
to a minimal set of scalar integrals called master integrals
and $(ii)$ their evaluation by means of differential equations 
in the masses or in the external kinematical invariants;
the differential equations are then solved using a proper basis of 
special functions, the harmonic polylogarithms 
\cite{eqdif,eqdif2,eqdif3,hpls,conrob2}.
Our aim is to present a simple introduction to this method.
Even though the latter has been used to do multi-loop calculations, 
it can also be used to reproduce standard one-loop results. 
Indeed, we are going to make use mostly of one-loop examples to describe the general
elements and ideas.
As we shall see, computations which are rather laborious with older 
techniques become much simpler with the differential equations and with the 
harmonic polylogarithms.
The plan of this note is the following.

In sec.~\ref{iniziale} we describe the tensor decomposition,
i.e. how the evaluation of tensor integrals, coming directly
from the application of Feynman rules, can be reduced
to that of scalar integrals. This step is well known to many
people and, strictly speaking, it does not belong to the method under 
discussion; it is included for completeness.

In sec.~\ref{indipendenti} we describe two widely-used schemes
to trasform the dependent scalar amplitudes generated with the
previous step into a smaller set of linearly-independent ones.
We present both one-loop and two-loop examples.

In sec.~\ref{diffids} we derive and solve the so-called
integration-by-parts identities, which allow to reduce
the independent amplitudes to a (much) smaller subset, 
the so-called master integrals.
The two main methods of solutions are discussed by means of
simple examples.

In sec.~\ref{methoddif} we describe the method of the differential
equations to analitically evaluate the master integrals.
We consider a couple of one-loop examples which exhibit many
of the general properties of the method.

In sec.~\ref{HPolylog} we overview the main ideas and results of the harmonic
polylogarith theory, including also the extension of the basis
funcion set to describe amplitudes with threshold at $s=4 m^2$.

\section{Feynman diagrams}
\label{iniziale}

The evaluation of virtual corrections to a cross section
begins with the application of Feynman rules to the relevant diagrams. 
Delicate points are typically the inclusion of the correct
multiplicity factors, the signs of fermion loops and, whenever
gauge interactions are present, a convenient gauge choice.
Nowadays this step can be done in an automated way \cite{feynarts}.

Let us consider for instance the top contribution to
Higgs production by gluon fusion, i.e. the process
\be
\label{gluonfus}
g+g\rightarrow H.
\ee 
The Feynman amplitude reads:
\be
{\cal M} \, = \, \epsilon_{\mu}(p_1) \, \epsilon_{\nu}(p_2) \, 
             \mathrm{T}^{\mu\nu}(p_1,p_2), 
\ee
where $\epsilon_{\mu}(p_1)$ and $\epsilon_{\nu}(p_2)$ are the polarizations of 
the gluons with momenta $p_1^2=0$ and $p_2^2=0$ and 
$\mathrm{T}^{\mu\nu}(p_1,p_2)$ is the following tensor:
\be
\mathrm{T}^{\mu\nu}(p_1,p_2) =
- \frac49 \,\,e^2\, N_c\, \frac{m_t}{2\, m_W}
\int \frac{d^n k}{(2\pi)^n} \, \mathrm{Tr} \left[ \,
 \gamma^{\mu} 
\frac{\hat{k}+\hat{p}_1+m_t}{(k+p_1)^2-m_t^2} 
\frac{\hat{k}-\hat{p}_2+m_t}{(k-p_2)^2-m_t^2}
 \gamma^{\nu} \frac{\hat{k}+m_t}{k^2-m_t^2} \right],
\ee
where $N_c$ is the color factor, $g$ the SU(2) coupling and $n$ the dimension 
of the space-time.
By evalutating the trace, the tensor comes out to be of the form:
\be
\mathrm{T}^{\mu\nu}(p_1,p_2) =
\int \frac{d^n k}{(2\pi)^n} \, 
\frac{N^{\mu\nu}(p_1,p_2,k)}
{ \left[(k+p_1)^2-m_t^2\right] \left[(k-p_2)^2-m_t^2\right] 
\left[k^2-m_t^2\right] }, 
\label{generalT}
\ee
where $N^{\mu\nu}(p_1,p_2,k)$ is a tensor depending on the external momenta
$p_1$ and $p_2$ as well as on the loop momentum $k$:
many possibile tensor structures are possible with 3 momenta,
\be
g^{\mu\nu},~~~~~
k^{\mu} k^{\nu},~~~~~ k^{\mu} p_1^{\nu},~~~~~ p_1^{\mu} k^{\nu},
~~~~~ k^{\mu} p_2^{\nu},
~~~~~ p_2^{\mu} k^{\nu},
~~~~~ p_1^{\mu} p_1^{\nu},
~~~~~ p_2^{\mu} p_2^{\nu},
~~~~~ p_1^{\mu} p_2^{\nu},
~~~\cdots 
\ee
The following reduction is convenient.
According to relativistic invariance, the tensor can be parametrized as 
\cite{noi4}
\bea
\mathrm{T}^{\mu\nu}(p_1,p_2) &=&
   p_1^{\mu} p_1^{\nu}\,\mathrm{T}_1(q^2)
 + p_2^{\mu} p_2^{\nu}\,\mathrm{T}_2(q^2)
 + p_1^{\mu} p_2^{\nu}\,\mathrm{T}_3(q^2)
 + p_2^{\mu} p_1^{\nu}\,\mathrm{T}_4(q^2)
 + g^{\mu \nu} (p_1\cdot p_2)\,\mathrm{T}_5(q^2) \nonumber \\
& + & \epsilon^{\mu \nu \rho \sigma} p_{1\rho} p_{2\sigma}\,\mathrm{T}_6(q^2),
\eea
where $q=p_1+p_2$ is the Higgs momentum and $\epsilon_{\mu \nu \rho \sigma}$
is the antisymmetric tensor.
The last form factor $T_6$ is related to parity violation of weak interactions.
Gauge invariance implies:
\be
p_1^{\mu} \, \mathrm{T}_{\mu\nu} = p_2^{\nu} \, \mathrm{T}_{\mu\nu} = 0,
\ee
which imply, in turn:
\be
\mathrm{T_1}~=~\mathrm{T_2}~=~0~~~~~~\mathrm{and}~~~~\mathrm{T_5}~=
-\mathrm{T_4}.
\ee
The above relations can be used as checks of the computation.
The form factors $T_i$ can be derived from the tensor $T_{\mu\nu}$ by means of 
projectors.
We have for instance:
\be
\mathrm{T}_5 = \frac{1}{(n-2)\, p_1\cdot p_2} \left[ g^{\mu\nu} -
\frac{p_2^{\mu} p_1^{\nu}}{p_1\cdot p_2} \right]
\mathrm{T}_{\mu\nu}.
\ee
By applying the projector to both sides of eq.~(\ref{generalT}) 
and taking it inside the loop integral, one obtains:
\be
\label{T5N5}
\mathrm{T}_5 = \int \frac{d^n k}{(2\pi)^n} 
\frac{N_5(k,p_1,p_2)}{ \left[(k+p_1)^2-m_t^2\right] 
\left[(k-p_2)^2-m_t^2\right] \left[k^2-m_t^2\right] },
\ee
where:
\be
N_5(k,p_1,p_2) = \frac{1}{(n-2)\, p_1\cdot p_2} \left[ g^{\mu\nu} -\frac{p_2^{\mu} p_1^{\nu}}
{p_1\cdot p_2} \right]
N_{\mu\nu}(k,p_1,p_2).
\ee
We have now a {\it scalar} numerator instead of a {\it tensor} one, depending 
only on invariants:
\be
N_5(k,p_1,p_2) = P(k^2,k\cdot p_1,k\cdot p_2),
\label{toinvar}
\ee
where $P$ is a polynomial:\footnote{
This is true is all {\it local} quantum field theories.}
\bea
P(k^2,k\cdot p_1,k\cdot p_2) 
&=& \sum_{l,r,s=0}^{n_{max}} \, a_{lrs}\,\left(k^2\right)^l 
\left(k\cdot p_1\right)^r \left(k\cdot p_2\right)^s
\nonumber \\
&=& a_{000} + a_{100} \, k^2 + a_{010} \, k\cdot p_1 +  a_{001} \, k\cdot p_2 
+ a_{200} \, \left( k^2 \right)^2 + a_{110} \, k^2 \, k\cdot p_1  + \cdots~~~~
\label{expansion}
\eea
and $a_{lrs}$ are known constants. $n_{max}$ is the maximum number 
of invariants and depends on the interaction; typically, in one-loop
computations, $n_{max}=1,\,2$.  
By using eqs.~(\ref{T5N5}), (\ref{toinvar}) and (\ref{expansion}), we obtain 
for the form factor $T_5$:
\be
T_5 \, = \, \sum_{l,r,s=0}^{n_{max}} \, a_{lrs} \int \frac{d^n k}{(2\pi)^n} 
\frac{\left(k^2 \right) ^l \left(k\cdot p_1\right)^r 
\left(k\cdot p_2\right)^s }
{  \left[(k+p_1)^2-m_t^2\right] \left[(k-p_2)^2-m_t^2\right] \left[k^2-m_t^2\right]  }.
\label{depscal}
\ee

\section{Independent amplitudes}
\label{indipendenti}

As we have seen, the calculations of the radiative corrections
to a generic process can be reduced to  the evaluation of  scalar integrals, 
that in a $ 2 \rightarrow 1$ process according to eq.~(\ref{depscal}) are of 
the form
\be
\int \frac{d^n k}{(2\pi)^n} 
\frac{\left(k^2 \right) ^l \left(k\cdot p_1\right)^r \left(k\cdot p_2\right)^s }{D_1 \, D_2 \, D_3},
\ee
where we have defined:
\bea
D_1 &=& k^2 - a,
\nonumber \\
D_2 &=& (k+p_1)^2 - a, 
\nonumber \\
D_3 &=& (k-p_2)^2 - a
\eea
with $a = m^2$  a generic mass squared.

The above amplitudes are linearly dependent on each other
and it is convenient to reduce them to a smaller set of
linearly independent ones.
We express the kinematical invariants in terms
of the denominators by means of the
following formulas --- the so-called rotation:
\bea
k^2 &=& D_1 + a,
\nonumber \\
k\cdot p_1 &=& \frac{1}{2} \left[ D_2 - D_1 \right],
\nonumber \\
k \cdot p_2 &=& \frac{1}{2} \left[ - D_3 + D_1 \right].
\label{rotations}
\eea
In general, in one-loop amplitudes, denominators form a basis for the
invariants.
There are two different schemes to implement the rotations 
given in eqs.~(\ref{rotations}):
\begin{enumerate}
\item
auxiliary diagram scheme;
\item
shift scheme.
\end{enumerate}

1) According to the first method, one uses eqs.~(\ref{rotations}) inside 
eq.~(\ref{depscal}), to obtain:
\be
T_5 \, = \, 
\sum_{l,r,s=0}^{n_{max}} b_{lrs} \, \mathrm{Topo}(1-l,1-r,1-s),
\ee
where $b_{lrs}$ are known constants and we have defined:
\be
\mathrm{Topo}(n_1,n_2,n_3) = \int \frac{d^n k}{(2\pi)^n} 
\frac{1}{ D_1^{n_1} \, D_2^{n_2} \, D_3^{n_3} }.
\ee
The amplitudes above constitute the linearly independent ones in this scheme.
Let us note that a propagator $D_i$ originally present in the denominator
may be cancelled by a term $D_i$, $D_i^2$, $D_i^3$, ... in the numerator.
In diagrammatric language, that means that internal line $i$ is shrunk 
to a point. The reduction to independent amplitudes then generates a pyramid
of subdiagrams of the original diagram, in which any subset of internal lines
is contracted to a point. 
For clariry's sake, one has to evaluate amplitudes of the form
\bea
&&
\frac{1}{D_1 \, D_2 \, D_3},
\nonumber\\
&&
\frac{1}{D_2 \, D_3},
~~~~~~\frac{D_1}{D_2 \, D_3},
~~~~~~\frac{D_1^2}{D_2 \, D_3},
~~~~\cdots
~~~~\frac{1}{D_1 \, D_3},
~~~~~~\frac{D_2}{D_1 \, D_3},
~~~~\cdots
~~~~\frac{1}{D_1 \, D_2},
~~~~~~\frac{D_3}{D_1 \, D_2},
~~~~\cdots
\nonumber\\
&& 
\frac{1}{D_1},
~~~~~~\frac{D_2}{D_1},
~~~~~~\frac{D_3}{D_1},
~~~~~~\frac{D_2^2}{D_1},
~~~~~~\frac{D_3^2}{D_1},
~~~~~~\frac{D_2 D_3}{D_1},
~~~\cdots
~~~\frac{1}{D_2},
~~~~~~\frac{D_1}{D_2},
~~~~~~\frac{D_3}{D_2},
~~~\cdots
~~~\frac{1}{D_3},
~~~~\cdots~~~
\eea

$2)$ Let us now consider the alternative scheme for the reduction to 
independent amplitudes.
The shift scheme makes use of the substitutions (\ref{rotations})
only when there is an effective cancellation of denominators.
In other words, the ``allowed'' substitutions are:
\bea
\frac{k^2}{D_1} &\rightarrow& 1 + \frac{a}{D_1},
\nonumber \\
\frac{k\cdot p_1}{D_1 \, D_2} &\rightarrow& \frac{1}{2} 
\left[ \frac{1}{D_1} - \frac{1}{D_2} \right],
\nonumber \\
\frac{k \cdot p_2}{D_1\, D_3} &\rightarrow& \frac{1}{2} 
\left[ - \frac{1}{D_1} + \frac{1}{D_3} \right].
\label{rotations2}
\eea
This way we are left with independent amplitudes of the following kinds:
\begin{enumerate}
\item
amplitudes containing only denominators: 
\bea 
\frac{1}{D_1 \, D_2 \,D_3}:~~~~~~~~~ && \mathrm{scalar~vertex,} \nonumber\\ 
\frac{1}{D_1 \,D_2 },~~~~~\frac{1}{D_1 \, D_3},~~~~~\frac{1}{D_2 \, D_3}: 
~~~~~~~~ && \mathrm{scalar~bubbles,} \nonumber\\ 
\frac{1}{D_1},~~~~~\frac{1}{D_2},~~~~\frac{1}{D_3}:~~~~~~~~ && 
\mathrm{scalar~tadpoles.}  
\eea 
The bubble diagrams are obtained shrinking any one of the internal lines
to a point, while tadpoles are obtained shrinking any pair
of internal lines;
\item 
bubble diagrams with irreducible numerators:\footnote{These amplitudes
as well as the following ones can be treated with a Passarino-Veltman
reduction; we choose to use the general method valid also in the
multi-loop case.}  \be 
\frac{k\cdot p_2}{D_1 \, D_2},~~~~~~~\frac{(k\cdot p_2)^2}{D_1 \, D_2}, 
~~~~~~~\frac{k\cdot p_1}{D_1 \, D_3}, ~~~~~~~\frac{(k\cdot p_1)^2}{D_1 \, D_3},
 ~~~\cdots, 
\ee 
and amplitudes which do not contain anymore $D_1$, i.e. the denominator
containing the loop momentum squared $k^2$: 
\be 
\frac{k^2}{D_2 \, D_3}, ~~~~~~~\frac{k\cdot p_1}{D_2 \, D_3}, 
~~~~~~~\frac{k\cdot p_2}{D_2 \, D_3}, ~~~~~~~~~\frac{(k^2)^2}{D_2 \, D_3},
~~~~~~~\frac{k^2~k\cdot p_1}{D_2 \, D_3}, ~~~~~~~\frac{k^2~k\cdot
p_2}{D_2 \, D_3}, ~~~\cdots
\label{toreduce}
\ee
No simplification is possible for the amplitudes in (\ref{toreduce})
as any cancellation in eq.~(\ref{rotations2}) is feasible only if $D_1$ is 
present.
We then make a shift of the loop momentum $k$ in order to
reproduce a denominator containing $k^2$, such as for instance:
\be
k \rightarrow k - p_1, 
\label{shift1}
\ee
so that 
\bea
D_2 &\rightarrow& D_1,
\nonumber\\
D_3 &\rightarrow& D_4 = (k-p_1-p_2)^2 - a.
\eea
The shift introduces therefore the new denominator $D_4$, not initially present
in the diagram.
Since $D_4$ contains both $p_1$ and $p_2$, one can express $k\cdot p_2$ in 
terms of $k\cdot p_1$ or {\it vice versa}. Let us take the first choice:
\be
\frac{k\cdot p_2}{D_1 D_4} \, = \, \frac{1}{2 D_4} - \frac{1}{2 D_1} - 
\frac{k\cdot p_1}{D_1 D_4}.
\ee
The amplitudes (\ref{toreduce}) are then transformed into amplitudes of the 
form:
\be
       \frac{1}{D_1 \, D_4},
~~~~~~~\frac{k \cdot p_1}{D_1 \, D_4},
~~~~~~~\frac{(k \cdot p_1)^2}{D_1 \, D_4},
~~~~~~~\frac{(k \cdot p_1)^3}{D_1 \, D_4},
~~~~\cdots
~~~~\mathrm{+~~~(tadpoles)};
\ee

\item Tadpoles, involving: 
\begin{itemize}
\item
''final'' amplitudes of the form:
\be
         \frac{k\cdot p_1}{D_1},
~~~~~~~~~\frac{k\cdot p_2}{D_1},
~~~~~~~~~\frac{(k\cdot p_1)^2}{D_1},
~~~~~~~~~\frac{(k\cdot p_2)^2}{D_1},
~~~~~~~~~\frac{k\cdot p_1 \, k\cdot p_2}{D_1}~;
~~~~~\cdots
\label{finaltad}
\ee
\item
reducible amplitudes of the form:
\bea
&&     \frac{k^2}{D_2},
~~~~~~~~~\frac{k\cdot p_1}{D_2},
~~~~~~~~~\frac{k\cdot p_2}{D_2},
~~~~~~~~~\frac{(k^2)^2}{D_2},
~~~~~~~~~\frac{k^2 \, k\cdot p_1}{D_2},
~~~~~~~~~\frac{k^2 \, k\cdot p_2}{D_2},
~~~~~\cdots
\nonumber\\
&&       \frac{k^2}{D_3},
~~~~~~~~~\frac{k\cdot p_1}{D_3},
~~~~~~~~~\frac{k\cdot p_2}{D_3},
~~~~~~~~~\frac{(k^2)^2}{D_3},
~~~~~~~~~\frac{k^2 \, k\cdot p_1}{D_3},
~~~~~~~~~\frac{k^2 \, k\cdot p_2}{D_3},
~~~~~\cdots
\eea
\end{itemize}
The amplitudes in the first line of the above expression are reduced 
by means of the shift in (\ref{shift1}), 
while for the second line we make the shift $k\rightarrow k+p_2$; 
the resulting amplitudes are of the form (\ref{finaltad}).
\end{enumerate}

The same paths of reduction to independent scalar amplitudes can be followed
for multiloop corrections. As a 2-loop example, let us now consider
the light-fermion correction to the process (\ref{gluonfus})
consisting of a ladder diagram. The latter describes a gluon pair converting 
into a light quark pair which converts in turn into a pair of W's or Z's 
annihilating finally into a Higgs boson.
The dependent scalar amplitudes are of the form:
\be
L = \int \frac{d^n k_1}{(2\pi)^n} \frac{d^n k_2}{(2\pi)^n}
\frac{P( k_1^2, k_2^2, k_1\cdot p_1, k_1\cdot p_2, k_2\cdot p_1, 
k_2\cdot p_2, k_1\cdot k_2)}
{ D_1\,D_2\,D_3\,D_4\,D_5\,D_6 },
\label{2Lamp}
\ee
where:
\bea
D_1 &=& k_1^2,
\nonumber\\
D_2 &=& (k_1+p_1)^2,
\nonumber\\
D_3 &=& (k_1-p_2)^2,
\nonumber\\
D_4 &=& k_2^2,
\nonumber\\
D_5 &=& (k_1+k_2+p_1)^2 - a,
\nonumber\\
D_6 &=& (k_1+k_2-p_2)^2 - a.
\eea
The conversion to independent amplitudes is not straightforward in 
this case because there are {\it six} denominators and {\it seven} invariants.
The denominators then do not form a basis for the invariants,
as it happened in the one-loop case.

The solution to this problem, in the auxiliary diagram scheme, 
is to construct an auxiliary diagram with an additional, fictitious
denominator linearly independent from the previous ones, such as for instance
\be
D_7 \, = \, (k_1+k_2)^2.
\ee
In diagrammatic language, we may say that we have ``opened'' the Higgs
vertex: the auxialiary diagram is a planar double box in a forward
configuration, i.e. with final momenta equal to the initial ones
$p_1$ and $p_2$.  
After the addition of the auxiliary denominator, the rotation is possible
by means of the formulas:
\bea
k_1^2 & = & D_1,
\nonumber\\
k_2^2 & = & D_4,
\nonumber\\
k_1\cdot p_1 & = & \frac{1}{2} (D_2-D_1),
\nonumber\\
k_1\cdot p_2 & = & \frac{1}{2} (-D_3+D_4),
\nonumber\\
k_2\cdot p_1 & = & \frac{1}{2} (D_1-D_2+D_5-D_7+a),
\nonumber\\
k_2\cdot p_2 & = & \frac{1}{2} (D_3-D_4-D_6+D_7-a),
\nonumber\\
k_1\cdot k_2 & = & \frac{1}{2} (D_7-D_1-D_4).
\eea
The numerator can then be expanded in powers of the denominators as:
\be
\label{num2L}
P( k_1^2, k_2^2, k_1\cdot p_1, k_1\cdot p_2, k_2\cdot p_1, k_2\cdot p_2, k_1
\cdot k_2) = \sum_{l_1,l_2,l_3,l_4,l_5,l_6,l_7=0}^{n_{max}}
c_{l_1 l_2 l_3 l_4 l_5 l_6 l_7}  
\, D_1^{l_1}\, D_2^{l_2} \,D_3^{l_3} \,D_4^{l_4} \,D_5^{l_5} \,D_6^{l_6} 
\,D_7^{l_7},~~
\ee
where $c_{l_1 l_2 l_3 l_4 l_5 l_6 l_7}$ are known constants.
By inserting the expansion (\ref{num2L}) in eq.~(\ref{2Lamp}), one obtains
independent amplitudes to be computeted, containing formally only denominators,
of the form:
\be
\mathrm{Topo}(n_1,n_2,n_3,n_4,n_5,n_6,n_7) = 
\int \frac{d^n k_1}{(2\pi)^n} \frac{d^n k_2}{(2\pi)^n}
\frac{1} { D_1^{n_1}\,D_2^{n_2}\,D_3^{n_3}\,D_4^{n_4}\,D_5^{n_5}\,D_6^{n_6}
\,D_7^{n_7} }
\ee
with $n_i\le 1$ for $i=1\ldots 6$ and $n_7\le 0$.
Note that the auxialiary denominator $D_7$ appears only in the numerator
while the standard denominators $D_i(i=1\ldots 6)$ appear both in the 
denominator and in the numerator.

As for the case of the auxiliary diagram scheme, also the shift method can be 
extended to the multi-loop case  in a straightforward way.

\section{Integration by parts identities}
\label{diffids}

The virtual corrections to a cross section 
involve, in general, the evaluation of a large number
of independent amplitudes --- in the case of massive 
two-loop computations hundreds if not thousands.
In the past all these amplitudes
had to be individually computed \cite{vanneerven}.
It is however possible to reduce by a large amount 
the number of amplitudes to be computed by using integral
identities \cite{firstibps}.
In sec.~\ref{subibp1} we discuss the derivation of the identities, while
in secs.~\ref{solve1} and \ref{solve2} we present two methods for their 
solution.
We work in the auxialiary diagram scheme; the discussion in the shift scheme
is completely analogous.

\subsection{Derivation of the identities}
\label{subibp1}

Let us begin with the simplest case, that of the one-loop tadpole.
According to the divergence theorem:
\be
\label{easyibp}
\int d^n k \frac{\partial}{\partial k_{\mu}} \, \frac{k_{\mu}}{D_1^{n_1}}  = 
\int_{S_{\infty}} ds^{\mu}  \frac{k_{\mu}}{D_1^{n_1}}   = 0,
\ee
where
\be
D_1 \, = \, k^2 - a,
\ee
$S_{\infty}$ is a sphere of infinite radius in momentum space and
$ds^{\mu}$ is a surphace element.
The flux integral actually vanishes only for $n_1>n/2$, but we will
analitically continue eq.~(\ref{easyibp}) to all the $(n,n_1)$ space.

By explicitly performining the derivative and re-expressing the result
in terms of independent amplitudes by means of the relation (see previous 
section)
\be
k^2 = D_1 + a, 
\ee
we obtain the following integration-by-parts (ibp) identity:
\be
\label{simplestibp}
\left(n-2n_1\right) \, T\left( n_1 \right) -2 a \, n_1 T\left(n_1+1\right) = 0,
\ee
where we have defined:
\be
T(n_1) = \int d^n k \frac{1}{D_1^{n_1}}. 
\ee

By introducing the identity operator $\mathrm{I}$ 
and the plus and minus operators,
\bea
\mathrm{I} \, T\left(n_1\right) &=& T\left(n_1\right),
\nonumber \\
1^{\pm} \, T\left(n_1\right) &=& T\left(n_1\pm 1\right),
\eea
the ibp identity can be written as:
\be
\left[\left(n-2n_1\right) \mathrm{I} - 2 a \, n_1 1^+ \right] 
T\left(n_1\right) ~=~ 0.
\ee

Let us now consider as a less trivial case: a bubble with one massive line,
\be
B(n_1,n_2) = \int d^n k \frac{1}{ D_1^{n_1} \, D_2^{n_2} },
\ee
where
\bea
D_1 & = & k^2 - a,
\nonumber\\
D_2 & = & (k+p)^2.
\eea
The integration-by parts identities are derived according to:
\be
\int d^n k \frac{\partial}{\partial k_{\mu}}~ \frac{v_{\mu}}{ D_1^{n_1} \, 
D_2^{n_2} } = 0,
\ee
where $v_{\mu}= k_{\mu}$ or $p_{\mu}$. 
We have therefore two identities for each set of indices $(n_1,n_2)$:
\bea
\label{ibp1}
&&  ( n-n_1-2 n_2 ) \mathrm{I} - n_1 \left[ a \, (1+x) + 2^- \right] 1^+ = 0, 
\\
\label{ibp2}
&&  (n_1-n_2) \mathrm{I} + n_1 \left[ a \, (1-x) - 2^- \right] 1^+ 
              + n_2 \left[ a \, (1+x) + 1^- \right] 2^+ = 0,
\eea
where $x=-p^2/a$ and all the operators are intended to be applied to 
$B(n_1,n_2)$.
Three different kinds of operators do appear in the identities:
\be
\mathrm{I},~~~~~~~i^+,~~~~~~~i^+ j^-,~~~~~~~~(i \ne j\, =\, 1,2)~. 
\ee
The generalitation to multi-loop multi-leg amplitudes is obvious. 
In the case for instance of the two-loop ladder diagram of the previous 
section, we have:
\be
\int d^n k_1 d^n k_2 \frac{\partial}{\partial k^{\mu}_j } 
\, \frac{v_{\mu}}
{ D_1^{n_1} \, D_2^{n_2} \, D_3^{n_3} \, D_4^{n_4} \, D_5^{n_5} \, D_6^{n_6} 
\, D_7^{n_7} } = 0,
\ee
where $j=1,2$ and $v~=~k_1,~k_2,~p_1,~p_2$ is any one of the loop or external 
momenta.
We have eight identities for any choice of the indices. Let us note that, in 
general, the identities are not all independent on each other.

\subsection{Symbolic solution}
\label{solve1}

Once the identities have been generated, the next step is to solve them
in a convenient way \cite{firstibps}. 
Let us begin with the tadpole identity (\ref{simplestibp}). 
We can solve it with respect to the amplitude with the greater index:
\be
T\left(n_1+1\right) = \frac{n-2n_1}{2 a \, n_1} T\left( n_1 \right).
\ee
If we assume for instance $T(1)$ to be known, we can determine from the above
equation $T(2),\,T(3),\,T(4),\,\cdots$, so that we can write:
\be
T(k) = a(k) \, T(1),
\ee 
where $a(k)$ is a known coefficient and $k$ is an integer.
We can also solve eq.~(\ref{simplestibp}) with respect to the amplitude with 
the smaller index:
\be
T\left(n_1\right) = \frac{2 a \, n_1}{n-2n_1} \, T\left( n_1+1 \right).
\ee
By setting $n_1=0$ we obtain $T\left(0\right)=0$, and hence 
$T\left(n_1\right)=0$ for
$n_1<0$, as well known from elementary quantum field theory computations.
The conclusion is that the tadpole topology has one master integral, which can 
be taken as $T(k)$ with $k$ a positive integer.

Let us now solve the bubble identities (\ref{ibp1}) and (\ref{ibp2}).
It is convenient to introduce the sum of the indices
\be
\Sigma \, = \, n_1 + n_2.
\ee
The plus operators $1^+$ and $2^+$ increase $\Sigma$ by one, while the identity
$\mathrm{I}$  and the plus-minus operators $1^+2^-$ and $2^+1^-$ keep 
$\Sigma$ unchanged. 
Let us assume that this topology has one master integral, which we take as 
$B(1,1)$,
having $\Sigma=2~$ --- this will be proved {\it a posteriori}.
A general amplitude, with $n_1\ge 1$ and $n_2 \ge 1$, has $\Sigma \ge 2$. That 
means we have to reduce $\Sigma$ by solving the above identities with respect 
to the plus operators. 
The first equation is solved with respect to $1^+$:
\be
1^+ = \frac{n-n_1-2n_2}{a n_1(1+x)}\,\mathrm{I} - \frac{1}{a(1+x)} \, 1^+ 2^-. 
\label{sol1}
\ee
With this equation we can shift the first index $n_1>1$ down to the value 
$n_1=1$.
Let us remark that it is impossible to go further because the coefficients
have $n_1$ in the denominator and then become singular.
Similarly, the second equation can be used to shift the second index $n_2$ down
to one:
\be
2^+ = \frac{n_2-n_1}{a n_2 (1+x)} \, \mathrm{I} - \frac{n_1(1-x)}{n_2(1+x)} 
\, 1^+      + \frac{n_1}{a n_2(1+x)} \, 1^+ 2^-  - \frac{1}{a (1+x)} 1^- 2^+.
\label{sol2}
\ee
Because of the presence of the minus operators, amplitudes with one of the 
indices equal to zero such as $B(1,0)$, $B(0,1)$, $B(2,0)$, etc.,
are encountered. These amplitudes have one of the internal lines
shrunk to a point and are therefore tadpoles, whose reduction has already been
discussed.
By recursively using eqs.~(\ref{sol1}) and (\ref{sol2}) we can reduce any 
amplitude
$B(n_1,n_2)$ with $n_1\ge 1$ and $n_2\ge 1$ to $B(1,1)$ $~+~$ (tadpoles):
\be
B(n_1,n_2) \, = \, c(n_1,n_2)~B(1,1) \, + \, d(n_1,n_2)\, T(1),
\ee
where $c(n_1,n_2)$ and $d(n_1,n_2)$ are known functions.
We have thus proved that the one-mass bubble has one master integral.

In some cases, amplitudes of a given topology can be reduced to subtopologies, 
i.e. to amplitudes with less internal lines. Let us consider as specific 
example a vertex diagram representing the annihilation of two massless 
particles with momenta $p_1$ and $p_2$ into a virtual particle with momentum 
$q=p_1+p_2$:
\be
V(n_1,n_2,n_3) = \int d^n k 
\frac{1}{ \left[ (k-p_2)^2 -a \right]^{n_1} \left[(k+p_1)^2\right]^{n_2} 
\left[k^2\right]^{n_3} }
\ee
with $p_1^2=0$ and $p_2^2=0$.
There is a massive line ``on a side'', connecting one of the massless particles
with the virtual one.
The following two ibp identities are easily derived:
\bea
&& (n-n_1-n_2-2n_3) \mathrm{I} - n_1 ( a + 3^- ) 1^+  - n_2 2^+ 3^- = 0,
\\
&& (-n_2+n_3) \mathrm{I} + n_1 (-xa - 2^- + 3^- ) 1^+  + n_2 2^+ 3^-  - 
n_3 3^+ 2^- = 0,
\eea
where $x=-q^2/a$.
By solving the first equation with respect to the $1^+$ operator and 
substituting the solution into the second equation, one obtains:
\be
      a \left[ (-n_2+n_3) - (n-n_1-n_2-2n_3) x \right] \mathrm{I} 
+ n_1 a \left[ - 2^-  +  (1+x) 3^- \right] 1^+ 
+ n_2 a(1+x) 2^+ 3^-
- n_3 a 3^+ 2^- = 0.
\ee
The above equation does not contain anymore plus operators, bringing unknown 
amplitudes, but only the identity and the plus-minus operators. By setting 
$n_1=n_2=n_3=1$ it is
immediantely seen that the basic amplitude $V(1,1,1)$ is expressed in terms of 
amplitudes
having one of the indices zero, i.e. of bubbles: we succeeded in the 
above-mentioned reduction.

\subsection{Laporta method}
\label{solve2}

This method has been originally introduced in \cite{laporta}
and has since then been widely used for the evaluation
of 2-loop 3-point and 4-point functions in a variety of mass 
and kinematical configurations \cite{eqdif3,massive1,conrob1,conrob2}.
The idea is that of replacing explicit values for the indices 
$n_i=\cdots -1,0,1\cdots$
in the ibp identities. This way a system of linear equations is generated,
whose unknowns are the amplitude themselves.
In the simple case of the tadpole, for instance, one generates a system of 
equations of the form:
\bea
2 a T(2) - (n-2) T(1) & = & 0,
\nonumber\\
4 a T(3) - (n-4) T(2) & = & 0,
\nonumber\\
6 a T(4) - (n-6) T(3) & = & 0,
\nonumber\\
\cdots &\cdots& \cdots
\nonumber\\
2 k \, a T(k+1) - (n-2 k) T(k) & = & 0.
\eea
The system is then solved with the method of elimination of variables of Gauss.
One has to decide which amplitudes have to be solved first. In the above 
example,  one could solve first for $T(k)$, then for $T(k-1)$, and so on.
In general, a good criterion is the following \cite{solve}:
\begin{itemize}
\item
We solve first for the amplitudes with the largest number of denominators.
More formally, we define the recursive parameter:
\be
\Sigma_1 = \sum_i \theta(n_i),
\ee
where the step function is defined as $\theta(u)=1$ if $u>0$ and zero 
otherwise, and we solve first for the amplitudes with the greatest $\Sigma_1$;
\item
Among the amplitudes with the same number of denominators, i.e. with the same
value of $\Sigma_1$, we solve first for those ones with the greatest sum of the
indices of the denominators,
\be
\Sigma_2 = \sum_i n_i \, \theta(n_i);
\ee
\item
finally, among the amplitudes with the same values for $\Sigma_1$ and 
$\Sigma_2$, we solve first for the amplitudes with the largest number of 
$D_i$ in the numerator -- in the shift scheme, that is the largest number of 
irreducible numerators:
\be
\Sigma_3 = \sum_i n_i \, \theta(-n_i).
\ee
\end{itemize}
Many amplitudes have, in general, the same values of 
$(\Sigma_1,\Sigma_2,\Sigma_3)$:  the choice of the amplitude can be random. 
Furthermore, a given amplitude appears, in general,  
in various equations: also the choice of the equation can be random.

According to Gauss method, we proceed with the progressive elimination of 
variables  till all the equations have been used. 
The amplitudes which remain on the r.h.s. at the end are the master integrals.
Within the scheme $(\Sigma_1,\Sigma_2,\Sigma_3)$ presented above, 
the master integrals typically involve amplitudes
with unitary denominators ($n_i=1$) and irreducible numerators 
($n_j=0,-1,-2\cdots$).
If we exchange $\Sigma_2$ with $\Sigma_3$, the master integrals
typically involve amplitudes with denominators squared.
In practise, one usually starts with a small linear system and looks at the 
master integrals coming from its solution. By enlarging the size of the system,
a smaller or equal set of master integrals is obtained. The idea of the method 
is that there is a ``critical mass'' of equations, above which a complete 
reduction to the master integrals occurs. The reason for this is that, by 
enlarging the system,  the number of equations grows faster than the 
number of unknowns \cite{eqdif3}.
The main virtue of this method is that it can be automated in a 
rather general way.

\section{The method of differential equations}
\label{methoddif}

Differential equations  in the masses or in the external kinematical invariants
offer a general method for the calculation of master integrals.
This method allows in principle to compute any loop amplitude which involves
more than one scale\footnote{Bubble diagrams, vertex diagrams with two external
particles on the light-cone, etc., having only massless propagators are then 
excluded. In all these cases, the differential equation gives only a 
dimensional, trivial information.}.
In sec.~(\ref{geneq}) we sketch the derivation of the differential equation for
the case of the one-mass bubble considered before,  while in 
secs.~(\ref{solveq1}) and (\ref{solveq2}) we describe the general method
to solve the equation.

\subsection{Generation of the equation}
\label{geneq}

Let us begin with perhaps the simplest possible example: 
the bubble with one massive and one massless line,
\be
B(p^2) = \int d^n \tilde{k} \frac{1}{ (k^2-a) \, (k+p)^2},
\ee
where
\be
d^n \tilde{k} \equiv a^{2-n/2} \frac{d^n k}{i \pi^{n/2}\Gamma(3-n/2)}. 
\ee
This diagram has a threshold at $p^2 = m^2$.
To obtain the differential equation, we take a derivative of the master 
integral with respect to the external invariant $p^2$ using the formula:\footnote{
We could derive with respect to the mass squared $a$ as well.}

\be
\frac{d}{d p^2} B(p^2) = \frac{1}{2 p^2} p_{\mu }\frac{\partial}{\partial p_{\mu}} B(p^2).
\ee
The partial derivative is taken inside the integral and produces various scalar
amplitudes, which are reduced to the master integral itself by means of the 
methods described in the previous sections.
The differential equation then closes on the master integral itself:
\be
\frac{d}{dx} B(x;\epsilon) = \left[ -\frac{1}{x}+\frac{1}{1+x} \right] \, 
B(x;\epsilon) +\epsilon \left[ \frac{1}{x}-\frac{2}{1+x} \right] \, 
B(x;\epsilon) +(1-\epsilon) \left[ \frac{1}{x}-\frac{1}{1+x} \right] \, 
T(\epsilon),
\label{bolla1m}
\ee
where 
\be
T(\epsilon) \, = \, \frac{1}{\epsilon} 
\ee
is the tadpole divided by $a$,
\be
\epsilon\, \equiv \, 2 - \frac{n}{2} 
\ee
and
\be
x\, \equiv \, -\frac{p^2}{a}.
\ee
We have included a minus sign in the definition of $x$ so that the (simpler) 
euclidean region $p^2<0$ corresponds to $x>0$.
The presence of the threshold in $p^2=a$ is reflected by the term $1/(1+x)$ in 
the differential equation.
Let us note that in the above derivation there is nothing specific about the 
one-loop case so the method extends trivially to the multiloop case.

\subsection{Initial conditions}
\label{solveq1}

In order to obtain a unique value for the master integral, an initial condition
has to be imposed to the general solution of the differential equation.
That means we have to know the master integral in a given kinematical point 
$x$.  Let us consider our example.
Since $B(x;\epsilon)$ is regular for $x\rightarrow 0$, it holds:
\be
\lim_{x \rightarrow 0} x \frac{d}{dx} B(x;\epsilon) = 0.
\label{condphys}
\ee
Multiplying both sides of eq.~(\ref{bolla1m}) by $x$, taking the limit 
$x\rightarrow 0$ and using eq.~(\ref{condphys}), one obtains:\footnote{In this 
simple case,  the value of the bubble for $x=0$ --- equivalent to $p=0$ ---  
can also be obtained with partial fractioning:
\be
B(x=0;\epsilon) \, = \, \int d^n \tilde{k} \frac{1}{(k^2-a) \, k^2} 
\, = \, \frac{1}{a}\int d^n \tilde{k} \frac{1}{k^2-a} - \frac{1}{a}\int d^n 
\tilde{k}\frac{1}{k^2} \, = \, \frac{1}{a}\int d^n \tilde{k} \frac{1}{k^2-a} \,
 = \, T(\epsilon).
\ee
The integral of $1/k^2$ vanishes because the integrand is scaleless.}
\be
\label{initcond}
B(x=0;\epsilon) \, = \, T(\epsilon).
\ee
We have thus obtained the initial condition by studying the master integral 
close to zero momentum and using the differential equaiton itself.

\subsection{Recursive solution in $\epsilon$}
\label{solveq2}

An efficient method to solve the differential equation for the master 
integral involves the $\epsilon$-expansion of the equation itself.
Eq.~(\ref{bolla1m}) is of the general form:
\be
\label{general1}
\frac{d}{dx} B(x;\epsilon) = A(x;\epsilon)\, B(x;\epsilon)+ \Omega(x;\epsilon),
\ee
where the coefficient of the unknown function is a polynomial of first order in
$\epsilon$:
\be
A(x;\epsilon) = A_0(x) + \epsilon A_1(x),
\ee
with
\bea
A_0(x) &=& - \frac{1}{x} + \frac{1}{1+x},
\nonumber\\
A_1(x) &=& \frac{1}{x} - \frac{2}{1+x}.
\eea
The main point is that $A(x;\epsilon)$ does not contain $1/\epsilon$ poles: 
this is true in general.
$\Omega(x;\epsilon)$ is the known term of the differential equation and is 
associated to the tadpole --- in general it is related to the subtopologies:
\be
\Omega(x;\epsilon) = \frac{1}{\epsilon}\Omega_{-1}(x) + \Omega_0(x) + 
\Omega_1(x) + \Omega_2(x) + \cdots,
\ee
where
\bea
\Omega_{-1}(x) &=& \frac{1}{x}-\frac{1}{1+x},
\nonumber\\
\Omega_{0}(x) &=& -\frac{1}{x}+\frac{1}{1+x},
\nonumber\\
\Omega_{1}(x) &=& 0,
\nonumber\\
\Omega_{2}(x) &=& 0,
\nonumber\\
\cdots &\cdots& \cdots
\eea
In our example $\Omega(x;\epsilon)$ has only two non-zero terms; 
in more complicated cases, $\Omega(x;\epsilon)$ contains higher-order poles
and has an infinite number of positive powers of $\epsilon$.
Let us now expand the (unknown) master integral in powers of $\epsilon$; since 
the known term contains at most a simple pole, we expect the same to be true 
for the MI:
\be
B(x;\epsilon) = \frac{1}{\epsilon} B_{-1}(x) + B_0(x) + \epsilon B_1(x) +
\epsilon^2 B_2(x) + \cdots 
\label{expansionMI}
\ee
Substituting the expansion (\ref{expansionMI}) in eq.~(\ref{general1}) 
and equating the coefficients of the powers of $\epsilon,$
we obtain a series of chained differential equations:
\bea
\frac{d}{dx} B_{-1}(x) &=& A_0(x) \, B_{-1}(x) \, + \, \Omega_{-1}(x),
\nonumber \\ 
\frac{d}{dx} B_{ 0}(x) &=& A_0(x) \, B_{ 0}(x) \, + \, A_1(x) \, B_{-1}(x) \, 
+ \, \Omega_{0}(x),
\nonumber \\ 
\frac{d}{dx} B_{ 1}(x) &=& A_0(x) \, B_{ 1}(x) \, + \, A_1(x) \, B_{ 0}(x) \, 
+ \, \Omega_{1}(x),
\nonumber \\
\cdots &\cdots& \cdots 
\nonumber \\ 
\frac{d}{dx} B_{ k}(x) &=& A_0(x) \, B_{ k}(x) \, + \, A_1(x) \, B_{k-1}(x) \, 
+ \, \Omega_{k}(x),
\nonumber \\
\cdots &\cdots& \cdots 
\eea
The first equation, for the coefficient $B_{-1}(x)$ of the simple pole, is the 
first one to be solved.
Once $B_{-1}(x)$ is known, we can insert its value in the second equation for 
$B_0(x)$ and solve for the latter function, and so on.
In more formal terms, we can redefine the known term as
\bea
\tilde{\Omega}_{-1}(x) 
&\equiv& \Omega_{-1}(x),
\nonumber\\
\tilde{\Omega}_{k}(x;B_{-1},\cdots,B_{k-1}) 
&\equiv& A_1(x) \, B_{k-1}(x) +\Omega_{k}(x)~~~~~~\mathrm{for}~~~~k \, \ge \,0,
\eea
and rewrite the system as:
\bea
\label{bigsys}
\frac{d}{dx} B_{-1}(x) &=& A_0(x) \, B_{-1}(x) \, + \, \tilde{\Omega}_{-1}(x),
\nonumber \\ 
\frac{d}{dx} B_{ 0}(x) &=& A_0(x) \, B_{ 0}(x) \, + \, 
\tilde{\Omega}_{0}(x;B_{-1}),
\nonumber \\ 
\frac{d}{dx} B_{ 1}(x) &=& A_0(x) \, B_{ 1}(x) \, + \, 
\tilde{\Omega}_{1}(x;B_{-1},B_0),
\nonumber \\
\cdots &\cdots& \cdots 
\nonumber \\ 
\frac{d}{dx} B_{ k}(x) &=& A_0(x) \, B_{ k}(x) \, + \, 
\tilde{\Omega}_{k}(x;B_{-1},\cdots,B_{k-1}),
\nonumber \\
\cdots &\cdots& \cdots 
\eea
The system (\ref{bigsys}) is solved with the method of variation of constants 
of Euler,  which we now summarize.
Let us first consider the associated homogeneous equation for $B_k(x)$, i.e. 
the equation obtained by dropping the known term:
\be
\frac{d}{dx} \omega(x) \, = \, A_0(x) \, \omega(x).
\ee
An important point is that the above equation is the same for any $k$.
This equation is solved with separation of variables:
\be
\omega(x) =  \exp \int A_0(x) \, dx. 
\ee
In our example,
\be
\frac{d}{d x}  \omega(x) \, = \, \left( -\frac{1}{x}+\frac{1}{1+x} \right) 
\omega(x),
\label{omog1m}
\ee
whose solution is:
\be
\omega(x) = \frac{1+x}{x} ,
\ee
where we have taken equal to unity the integration constant.

The solution of the original, non-homogeneous equation is given by the 
following integral:
\be
B_{k}(x) = \omega(x) \int^x K(x') \, 
\tilde{\Omega}_{k}(x';B_{-1},\cdots,B_{k-1}) \, dx',
\ee
where the kernel $K$ is just the inverse of the homogeneous solution:
\be
K(x) \equiv \frac{1}{\omega(x)}~.
\ee
The simple pole of the one-mass bubble for instance is given by:
\be
B_{-1} = c \frac{1+x}{x} +\frac{1+x}{x} \int_0^x dx' \frac{x'}{1+x'} 
\left(\frac{1}{x'}-\frac{1}{1+x'}\right)
=\frac{c-1}{x} + c = 1,
\ee
where in the last member we have imposed the initial condition 
(\ref{initcond}).

As another example, let us consider the bubble with two equal masses:
\be
{\cal B}(x;\epsilon) = \int d^n \tilde{k} \frac{1}{ \left[k^2-a\right] \, \left[(k+p)^2-a\right] }.
\ee
This diagram has a threshold in $p^2 =  4 m^2$.
The differential equation reads:
\be
\frac{d}{dx} {\cal B}(x;\epsilon) = 
  \left[ -\frac{1}{2 x}+\frac{1}{2(4+x)} \right] \,   {\cal B}(x;\epsilon)
- \frac{\epsilon}{4+x} \, {\cal B}(x;\epsilon)
+(1-\epsilon) \left[ \frac{1}{2 x}-\frac{1}{2(4+x)} \right] \, T(\epsilon).
\label{bolla2m}
\ee
The solution of the associated homogeneous equation in four dimensions,
\be
\frac{d}{dx} \phi (x) = 
\left[ -\frac{1}{2 x}+\frac{1}{2(4+x)} \right] \, \phi(x),
\ee
is:
\be
\phi(x)  = \sqrt{ \frac{4+x}{x} }.
\ee
A first difference with respect to the one-mass case is that 
the term $1/(1+x)$ is replaced by the term $1/(4+x)$,
as a consequence of the threshold in $p^2=4m^2$ instead of in $p^2=m^2$.
Another less trivial difference is that semi-integer coefficients appear in the
homogeneous differential equation, leading to square roots in the solution 
$\phi(x)$.

\section{Harmonic Polylogarithms}
\label{HPolylog}
Let us write explicitly the formal solution of the set of differential 
equations considered in the previous section:
\bea
B_{-1}(x) &=& \omega(x) \int^x dx' K(x') \, \Omega_{-1}(x'),  
\nonumber\\
B_0(x) &=& \omega(x) \int^x dx' A_1(x') \int^{x'} dx'' K(x'') \, 
\Omega_{-1}(x'') +\omega(x) \int^x dx' K(x') \, \Omega_0(x'),
\nonumber\\
B_1(x) &=& \omega(x) \int^x dx' A_1(x') \int^{x'} dx'' A_1(x'') \int^{x''} dx''' 
K(x''') \, \Omega_{-1}(x''')
\nonumber\\
       &+& \omega(x) \int^x dx' A_1(x')\int^{x'} dx'' K(x'') \, \Omega_0(x'')
        + \omega(x) \int^x dx' K(x') \, \Omega_{1}(x'),
\nonumber\\
\cdots &\cdots& \cdots
\eea
As is clearly seen from the above expressions, the solutions of the 
differential equations involve repeated integrations of products of the kernel 
$K(x)$ and of coefficients of the differential equation itself: 
$A_1(x)$, $\Omega_{-1}(x)$,  $\Omega_0(x)$, etc.
Natural representations of the solutions seem therefore repeated integrations
of $K(x)$ and of the elementary functions entering the differential equation under study.
The idea behind the Harmonic Polylogarithms (HPLs) is simply that of giving a 
name to such repeated integrations \cite{hpls}.
For the one-mass bubble, for instance, it is natural to define:\footnote{
The function $1/(1-x)$ is introduced for the closure under the transformation
$x\rightarrow -x$.}
\bea
g(-1;x)&=&\frac{1}{1+x},
\nonumber\\
g(0;x)&=&\frac{1}{x},
\nonumber\\
g(1;x)&=&\frac{1}{1-x}.
\eea
The harmonic polylogarithms of weight one are defined as integrals
of the above functions:
\bea
H(-1;x)&=&\int_0^x \frac{dx'}{1+x'} = \log(1+x),
\nonumber\\
H(0;x)&=& \int_1^x \frac{dx'}{x'} = \log(x),
\nonumber\\
H(1;x)&=& \int_0^x \frac{dx'}{1-x'} = -\log(1-x).
\eea
Note the slight asymmetry in the lower limit of integration
of $H(0;x)$ related to non-integrable singularity of $1/x$ in $x=0$.
Harmonic polylogarithms of higher weight $w$ have the following 
integral recursive definition:
\be
H(a,\vec{w};x)~=~\int_0^x g(a;x')\, H(\vec{w};x') dx'
\ee
for $(a,\vec{w})\ne (0,\vec{0}_w)$ and
\be
H(\vec{0}_w;x)=\frac{1}{w!} \, \log^w (x).
\ee
The index $a$ takes the values $0,\pm 1$ and $\vec{w}$ is a string
of $w$ indices, each one taking the values $0,\pm 1$. 
The vector $\vec{0}_w$ is a string of $w$ zeroes. 

Using the above basis, the bubble with one mass reads:
\bea
B_{-1}&=&1,
\nonumber\\
B_0&=& 2 - \left( 1 + \frac{1}{x} \right) H(-1;x),
\nonumber\\
B_1&=& 4 - \left( 1 + \frac{1}{x} \right) 
\left[ 2 H(-1;x) + H(0,-1;x) - 2H(-1,-1;x) \right],
\nonumber\\
\cdots &\cdots& \cdots
\eea
Higher order terms in the $\epsilon$-expansion involve, as expected, 
harmonic polylogarithms of higher weight.

As far as the bubble with two masses is concerned, the above function set
is not sufficient\footnote{It is possible to solve the differential equation
with the ordinary harmonic polylogarithms by means of the change of variable
$x=(1-z)^2/z$, which eliminates the square roots.}. 
We add to the basis the functions \cite{conrob2}:
\bea
f(-4;x)&=&\frac{1}{4+x},
\nonumber\\
f(4;x)&=&\frac{1}{4-x},
\nonumber\\
f(-r;x)&=&\frac{1}{\sqrt{x(4+x)}},
\nonumber\\
f(r;x)&=&\frac{1}{\sqrt{x(4-x)}}.
\eea
The related harmonic polylogarithms of weigth $w\ge 1$ are defined analogously 
to the standard ones. By using this extended special function set, the two-mass
bubble reads:
\bea
B_{-1}&=&1,
\nonumber\\
B_0&=& 2 -\sqrt{\frac{x+4}{4}}H(-r;x),
\nonumber\\
B_1&=& 4 -\sqrt{\frac{x+4}{4}} \left[ 2 H(-r;x) - H(-4,-r;x) \right],
\nonumber\\
\cdots &\cdots& \cdots
\nonumber
\eea
The lesson is that loop diagrams are represented by complicated, special functions 
because they involve repeated integrations of simple basic functions.
In other words, the complexity of the results originates {\it solely} from the
repeated integrations.

\vskip 1truecm

\centerline{\bf Acknowledgements}

\vskip 0.5truecm

I would like to thank G. Degrassi for discussions.

\end{document}